\documentclass[conference,10pt]{IEEEtran}
\IEEEoverridecommandlockouts
\usepackage{graphicx}
\usepackage{times}
\usepackage[hyphens]{url}
\usepackage{caption}
\usepackage{subcaption}
\usepackage{cite}

\begin{document}

\title{Stealthy Malware Traffic \\ -- Not as Innocent as It Looks}

\author{Xingsi Zhong, \textit{Student Member, IEEE}, Yu Fu, Lu Yu, Richard Brooks,\\ \textit{Senior Member, IEEE} and G. Kumar Venayagamoorthy, \textit{Senior Member, IEEE}\\
Real-Time Power and Intelligent Systems Laboratory\\
Holcombe Department of Electrical and Computer Engineering\\
Clemson University, Clemson, SC, 29634, USA\\
Email: xingsiz@g.clemson.edu, fu2@g.clemson.edu,lyu@g.clemson.edu, \\rrb@acm.org and gkumar@ieee.org\\
\thanks{
This work was supported by the U.S. National Science Foundation, under Grant IIP \# 1312260. Any opinions, findings, and conclusions or recommendations expressed in this work are those of the authors and do not reflect the views of the National Science Foundation.}
}

\maketitle
\thispagestyle{empty}

\begin{abstract}
Malware is constantly evolving. Although existing countermeasures have success in malware detection, corresponding counter-countermeasures are always emerging. In this study, a counter-countermeasure that avoids network-based detection approaches by camouflaging malicious traffic as an innocuous protocol is presented. The approach includes two steps: Traffic format transformation and side-channel massage (SCM). Format-transforming encryption (FTE) translates protocol syntax to mimic another innocuous protocol while SCM obscures traffic side-channels. The proposed approach is illustrated by transforming Zeus botnet (Zbot) Command and Control (C\&C) traffic into smart grid Phasor Measurement Unit (PMU) data. The experimental results show that the transformed traffic is identified by Wireshark as synchrophasor protocol, and the transformed protocol fools current side-channel attacks. Moreover, it is shown that a real smart grid Phasor Data Concentrator (PDC) accepts the false PMU data. 
\end{abstract}

\section{Introduction}
Malware authors conceal Command and Control (C\&C) traffic\footnote{In the context of cybersecurity, attackers use C\&C infrastructure to issue C\&C instructions to their victims through network.} by masquerading as legitimate web browsing activity (HTTP or encrypted HTTPS)~\cite{gu2008botminer} or Internet relay chat (IRC).  Twitter and DNS queries have also been used to hide C\&C communication channels~\cite{morto}. Detection techniques based on reverse engineering, side-channel analysis, etc., have emerged~\cite{binsalleeh2010analysis, lu2011botnet}. Reverse engineering of malware develops network signatures for malicious traffic filtering~\cite{snort}. Many bots use cryptography or obfuscation to disrupt signature-based detection~\cite{apvrille2011cryptography}, which can be foiled by blocking all unknown encrypted traffic~\cite{china2012mike}. However, this has two drawbacks: (1) discouraging the use of cryptography makes traffic less secure and (2) any false positives disrupt legitimate traffic producing denial of service.
Other research uses traffic analysis for detection. Chen~\cite{lu2012network} presents timing side-channel analysis for Zbot detection using hidden Markov model (HMM) and probabilistic context-free grammar (PCFG). Passive DNS traffic analysis system for detecting and tracking malicious flux networks of a botnet is proposed in~\cite{perdisci2012early}. 
Other traffic side-channels include packet size~\cite{zhong2015twoPMU}.

In this paper, a \textit{counter-countermeasure} that obscures the features used by malware traffic detection tools is presented. The idea is to camouflage malware traffic as an innocuous application protocol using format-transforming encryption (FTE)~\cite{Dyer:2013:PMM:2508859.2516657} and side-channel massage (SCM). FTE converts traffic flow to another application protocol using a regular expression describing the target protocol. This thwarts existing detection approaches because they detect that the infected machines are using IRC, HTTP/HTTPS or DNS to communicate. Additionally, the FTE output is processed using SCM to modify side-channel features, making the system resistant to side-channel analysis. The motivation of this study is to reveal the weakness of existing traffic detection techniques and show that more sophisticated malware traffic detection solutions are needed. 

This counter-countermeasure is illustrated by camouflaging Zbot traffic. Zeus malware is notorious for stealing on-line financial information. The source code of Zeus is on Github~\cite{zeussrc}. In the experiment presented here, the Phasor Measurement Unit (PMU) communication protocol is the target protocol for the following reasons: (1) 
PMU traffic is 
in clear text~\cite{stewart2010synchrophasor, beasley2014survey}, and resistant to 
countermeasures that block encrypted traffic; (2) as early as 2009, the possibility of a power-grid botnet consisting of buggy smart meters was raised~\cite{smbot}; and (3) the synchrophasor protocol used for PMU communication 
is regular, simple and easy to imitate ``in its entirety''. The feasibility of converting botnet traffic to PMU traffic is explored in this study. Mimicking is done at the application layer of the protocol stack, which simplifies the imitation process and reduces the risk of errors. Experimental results show that the counterfeit PMU can talk with the real server without being detected. Since the same traffic can be processed to recover the C\&C traffic, the botnet communications are effectively concealed.

The remainder of the paper is organized as follows. Section~\ref{sec:rw} and Section~\ref{sec:background} provide related study and background information. 
The traffic detection counter-countermeasure is explained in Section~\ref{sec:camouflage}. In Section~\ref{sec:exp}, An example to demonstrate the utility of the approach is provided. Conclusions and suggestions for future study are given in Section~\ref{sec:candf}.

\section{Related Work}\label{sec:rw}
Malware uses traffic camouflaging to hide network traffic. Spyware Taidoor, a banking trojan discovered in 2008, used a Yahoo blog as a botnet C\&C server \cite{taidoor}, the malicious traffic was camouflaged as HTTP requests to blogs. According to Kaspersky Lab \cite{malwaretor}, ChewBacca and evolved Zeus use the Tor network to hide communications. Pushdo malware \cite{pushdo} included domain names of large companies or famous educational institutions in the C\&C domain list, which made network traffic analysis more difficult. The Morto Trojan \cite{morto} sent false DNS requests to a DNS server, which was a C\&C server, and the exchanged message was obfuscated by a simple Base64 encoding. The proposed method can transform arbitrary traffic to a target protocol, so that malware detection techniques looking for features of the original protocol won't work.

Besides hiding malware traffic, traffic camouflage can also be used to evade surveillance and censorship. The Onion Router (Tor) \cite{reed1998anonymous} provides an infrastructure for anonymous communication over a public network. The idea is to re-route network traffic through several browser-based short-lived proxies (called relay nodes) to evade surveillance and censorship \cite{fifield2012evading}. However, as Tor becomes well-known, some unpublished relay nodes (bridges) are blocked in authoritarian countries \cite{winter2012china}. 

To counter that, obfsproxy \cite{obfsproxy} is developed to circumvent censorship by camouflaging the Tor traffic between client and bridge. It supports multiple protocols, called pluggable transports, which specify how traffic is transformed. ScrambleSuit \cite{winter2013scramblesuit} is a polymorphic network protocol to obfuscate the transported application data to defend against active probing and protocol fingerprinting. SkypeMorph \cite{mohajeri2012skypemorph} is a Tor pluggable transport to reshape Tor packets to resemble Skype calls. StegoTorus \cite{weinberg2012stegotorus} first uses chopping to change packet sizes and timing information, and then uses steganography to disguise Tor traffic as a message in an innocuous cover protocol, such as HTTP. FTE \cite{Dyer:2013:PMM:2508859.2516657} evades regular-based deep packet inspection (DPI) technologies by transforming Tor traffic into a predefined format. Although FTE is effective in mimicking the contents of packets, it does not mimic timing information, which is vulnerable to side-channel analysis. The proposed SCM method `massages' the timing side-channel to ameliorate FTE.



\section{Background}\label{sec:background}
 	\subsection{Malware Traffic Detection}
Network-based malware detection techniques 
examine the behavior of underlying malicious activities 
in the network traffic. 
Early study on network-based detection inspect the network packets for known botnet signatures. 
DPI technologies are usually used~\cite{botdpi}. 
However, signature-based detection suffers from computational complexity and can be easily foiled by encryption and obfuscation. 

Other detection approaches include anomaly~\cite{binkley2006algorithm} detection, 
and traffic analysis~\cite{lu2012network,perdisci2012early}.  Although anomaly detection might detect unknown malware, it usually has high false positive rates (FPRs). 
Traffic analysis exploits artifacts.  Many properties of network traffic (packet size, power consumption, timing, electromagnetic radiation, etc.) remain observable after encryption and reveal valuable information. Attacks based on these properties are known as side-channel attacks. 
A timing side-channel attack is presented in~\cite{ryanthesis}, which breaks the anonymity of Tor network by matching packet timing data from any two Tor users to a model.  The comparison gives a statistical likelihood that the two systems are actually communicating over Tor. 
In~\cite{Zhong:2015:CISRC, zhong2015twoPMU}, it is demonstrated that inter-packet delays and packet size can differentiate between packet streams coming from different PMUs. 


Chen et al.~\cite{lu2012network} show Zbot traffic can be identified via timing side-channel analysis. The timing pattern in network traffic is represented by a deterministic hidden Markov model (HMM), which can be constructed from the collected inter-packet delays using the zero-knowledge HMM inference algorithm~\cite{schwier2009zero,yu2013inferring,chenmetric}. Zbot traffic detection is performed by comparing the constructed HMMs of the target traffic and Zbot traffic. This method is leveraged in this study to build the HMM used for SCM.

 	\subsection{Format-Transforming Encryption}
A network protocol is a set of rules for communication. 
An implicit premise most of DPI is that regular expressions are sufficient for identifying protocols~\cite{Dyer:2013:PMM:2508859.2516657}.  Tools, such as Snort \cite{snort}, Bro \cite{bro} and L7-filter \cite{l7filter}, use regular expressions to identify network protocols.

Format-Transforming Encryption (FTE) \cite{Dyer:2013:PMM:2508859.2516657} is developed to evade regular expression based protocol identification.  It extends conventional symmetric encryption 
by formatting the ciphertext.  Arbitrary application-layer network traffic can be transformed into a target protocol using FTE.  FTE is used to evade Internet censorship and surveillance, because the original protocols are obfuscated.

Although FTE is effective in hiding network protocols, it can be a challenge to write the regular expression for the target protocol. First, there is no automated software or procedures to extract regular expressions from protocols. Second, a regular expression that fully describes a protocol is usually complex. Although FTE can evade syntax-based protocol identification, it is vulnerable to side-channel analysis. Features like packet size, inter-packet delays etc., can be used for protocol identification. In this paper, a SCM method is developed to obscure side-channel features and augment FTE. 


The implementation of FTE includes LibFTE and FTEproxy. LibFTE \cite{libfte} is a python library to transform string to ciphertext. FTEproxy \cite{fteproxy} is a Tor \cite{Tor} pluggable transport layer to resist keyword filtering, censorship and discriminatory routing policies. LibFTE (version 0.1.0) \cite{libfte} is used in this study.

 	\subsection{Zeus Botnet}
Botnets are groups of compromised computers that botmasters (botherders) use to launch attacks over the Internet. As one of the most famous botnets, Zeus botnet is a malicious network that steals banking information with keystroke logging and screen capture \cite{lu2012network}. It can also be used in other types of data or identity theft \cite{zeus}. Since its first appearance in 2007, Zeus has been widespread and infecting Windows machines through drive-by downloads and phishing schemes. Sold on the black market, Zeus allows technically unsophisticated users to carry out cybercrimes. According to~\cite{russian} Trend Micro, Zeus (2.0.8.9) source code can cost \$400-500 in the Russian market. In 2011, the source code of Zeus was leaked on several underground forums over the Internet. A security researcher compiled the code and confirmed it worked `like a charm' \cite{zeusleaked}. 

\begin{figure}
\begin{center}
\includegraphics[width=.45\textwidth]{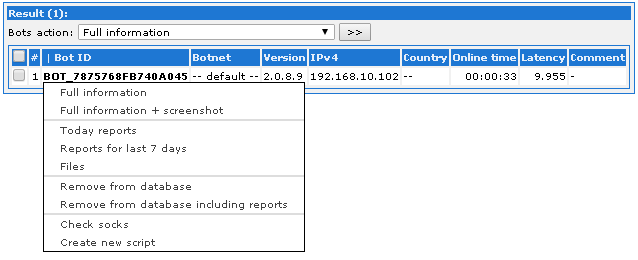}
\caption{Control panel of Zeus C \& C server}
\label{zeus-cp}
\end{center}
\end{figure}

\begin{figure}
\begin{center}
\includegraphics[width=.45\textwidth]{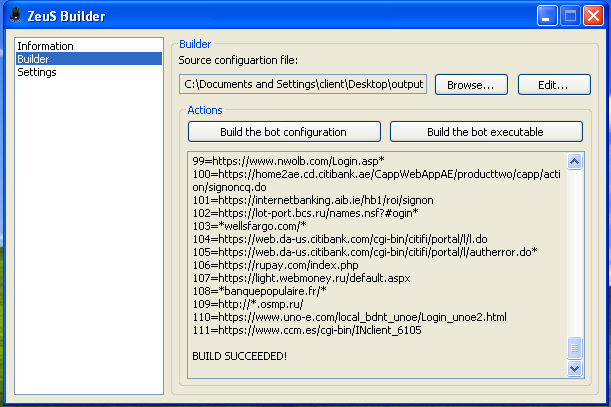}
\caption{Zeus builder to generate executables to infect bots}
\label{zeus-builder}
\end{center}
\end{figure}

Zbot (version 2.0.8.9) was retrieved from Github \cite{zeussrc} for experimentation. It contains code to build a Zeus Command and Control (C\&C) server and generate executables to infect victim machines. A Zeus C\&C server is a PHP server with mySQL and Apache. Every time a bot joins the network, the botmaster can monitor and control it through the control panel. Figure \ref{zeus-cp} shows the control panel of Zeus C\&C server. To generate executables to infect bots, Zeus has a builder tool, where you can specify the configuration directory file and C\&C server address. Figure \ref{zeus-builder} shows the Zeus builder tool GUI.

  \subsection{PMU in Smart Grid}
\begin{figure}[h]
\begin{center}
\includegraphics[width=2.5in]{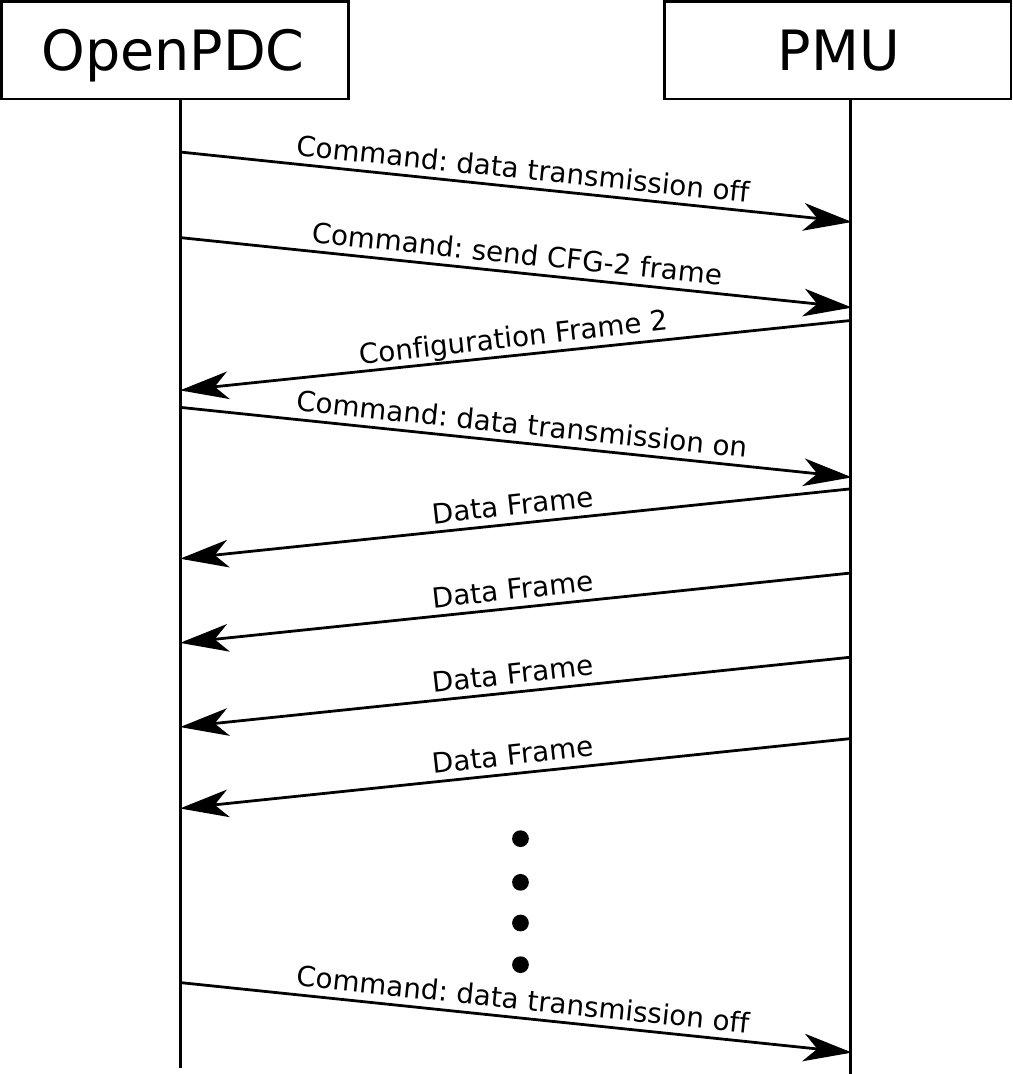}
\caption{Observed sequence diagram of Synchrophasor protocol in application layer}
\label{SynchrophasorSTDiagram}
\end{center}
\end{figure}

PMUs measure bus voltages, line currents, and bus frequencies in transmission systems in real-time. Each measurement is tagged with a global positioning system (GPS) time stamp \cite{Zhong:2015:CISRC, zhong2015twoPMU}.  Measurement data is sent to Phasor Data Concentrators (PDCs) via TCP/IP for further processing \cite{beasley2014survey}.  The communications between PMUs and PDCs use the synchrophasor protocol. In this study, open source software OpenPDC~\cite{openPDC} is used. A typical sequence diagram of communications between OpenPDC and one PMU in the application layer is shown in Figure \ref{SynchrophasorSTDiagram}. OpenPDC starts a new session by sending a command asking the PMU to stop current data transmission, followed by a request for a configuration frame, which describes the format of PMU measurement packets used for this transmission. After receiving the configuration frame from PMU, OpenPDC sends the PMU a start command. Once the measurement data transmission starts, it will not stop until a stop command is received or the connection is lost. OpenPDC acknowledges each measurement data packet it receives.

\section{Traffic Camouflage}\label{sec:camouflage}
The counter-countermeasure consists of 2 steps: (1) Zbot traffic is transformed using FTE to hide the source protocol; (2) the output of FTE is improved with SCM to modify inter-packet delays to evade side-channel analysis. 


 	\subsection{FTE}
The input of FTE includes the message to encrypt and the regular expression describing the target protocol. The output ciphertext matches the input regular expression. There is a parameter, called $fixed\_slice$, which refers to the number of bytes that FTE receiver-side should decrypt of an incoming stream. 
For example, the target regular expression is `$\wedge (a|b)+\$$', which matches an arbitrary long string only comprising $a$ or $b$. Given a message $hello world$ and $fixed\_slice=512$, a string of $aaaaaaaabababaab...bababbabaaababab$ containing 512 characters is obtained.


The goal is to transform Zbot traffic into a valid PMU data stream. 
PMU data is collected to infer a regular expression for synchrophasor protocol.  Table \ref{pmu-value} shows the structure of a PMU data packet~\cite{6111222}, and the number of observed values for all fields.  The extracted regular expression that fully describes observed PMU traffic is very complex. The transformation using this regular expression requires excessive computational resources.


\begin{table}[h]
\centering
\caption{PMU packet fields in application layer}
\label{pmu-value}
\begin{tabular}{|c|c|c|c|}
\hline
\begin{tabular}[c]{@{}c@{}}Byte \\ Poisition\end{tabular} & \begin{tabular}[c]{@{}c@{}}Length \\ in Bytes\end{tabular} & \begin{tabular}[c]{@{}c@{}}Parameter \\ Name\end{tabular}                 & \begin{tabular}[c]{@{}c@{}}Number of \\ observed \\ values\end{tabular} \\ \hline
0 - 1                                                   & 2                                                          & \begin{tabular}[c]{@{}c@{}}Synchroni-\\ zation word\end{tabular}          & 1                                                                       \\ \hline
2 - 3                                                   & 2                                                          & Framesize                                                                 & 1                                                                       \\ \hline
4 - 5                                                   & 2                                                          & \begin{tabular}[c]{@{}c@{}}PMU/DC \\ ID number\end{tabular}               & 1                                                                       \\ \hline
6 - 9                                                   & 4                                                          & \begin{tabular}[c]{@{}c@{}}SOC time \\ stamp (UTC)\end{tabular}           & 3958                                                                    \\ \hline
10                                                        & 1                                                          & \begin{tabular}[c]{@{}c@{}}Time quality \\ flag\end{tabular}              & 4                                                                       \\ \hline
11 - 13                                                   & 3                                                          & \begin{tabular}[c]{@{}c@{}}Fraction of \\ second (raw)\end{tabular}       & 30                                                                      \\ \hline
14 - 15                                                   & 2                                                          & Flags                                                                     & 4                                                                       \\ \hline
16 - 271                                                  & \begin{tabular}[c]{@{}c@{}}8 \\ each\end{tabular}          & \begin{tabular}[c]{@{}c@{}}Phasor \#1 \\ - \#32\end{tabular}              & \begin{tabular}[c]{@{}c@{}}118713 \\ each\end{tabular}                  \\ \hline
272 - 275                                                 & 4                                                          & \begin{tabular}[c]{@{}c@{}}Actual \\ frequency \\ value\end{tabular}      & 52                                                                      \\ \hline
276 - 279                                                 & 4                                                          & \begin{tabular}[c]{@{}c@{}}Rate of \\ change of \\ frequency\end{tabular} & 7573                                                                    \\ \hline
280 - 281                                                 & 2                                                          & \begin{tabular}[c]{@{}c@{}}Digital status \\ words\end{tabular}           & 1                                                                       \\ \hline
282 - 283                                                   & 2                                                          & PMU checksum                                                              & 54887                                                                   \\ \hline
\end{tabular}
\end{table}

In this study, the FTE performance if improved by mapping observed PMU traffic directly to a regular expression. The Zbot traffic is first transformed with FTE using an example regular expression `$\wedge[0-9a-f]+\$$'. The output string consists of a sequence of hexadecimal symbols. There are 16 possibilities for each character position. Each character position corresponds to one PMU packet field. For each PMU packet field (Figure \ref{pmu-value}), the observed values of that field are divided into 16 groups and assign each group a unique symbol. Based on the predefined mapping relation from hexadecimal symbols to groups of value, for each character in the output string of FTE, a value is randomly picked from the related group. 

Note that the previous procedure only applies to fields with at least 16 possible values. For fields with less than 16 possible values, a random value from the observation is used. 
For example, suppose the FTE output is `a10b5d7...'. The first character `a' corresponds to group number 10. So it can be replaced by a random value in the 10th group of the first field of PMU packet, which is `SOC time stamp (UTC)'. 

There are several advantages of the proposed mapping technique: (1) The input regular expression is straightforward.  There is no need to extract regular expressions from target protocols. (2) Since regular expression extraction is not needed, the process can be easily automated. (3) The proposed mapping uses observed data, so the output is closer to real PMU traffic than traditional FTE. (4) The throughput can be increased by adjusting the number of symbols used to represent one protocol field.


 \subsection{Timing Side-Channel Massage (SCM)}\label{subsec:scm}

Timing SCM modifies inter-packet delays to match the target protocol to evade side-channel analysis.
Given collected inter-packet delays of the target protocol, a deterministic HMM representing the timing pattern is constructed.  A HMM is a statistical Markov model in which the modeled system is a Markov process with unobserved (hidden) states.  The standard HMM in~\cite{rabiner1989tutorial} has two sets of random processes: states and outputs (observations).  In a deterministic HMM, there is only a single set of random processes.  In this model, the state transition labels are uniquely associated with an output alphabet.

\begin{figure}[h]
\begin{center}
\includegraphics[width=3.0in]{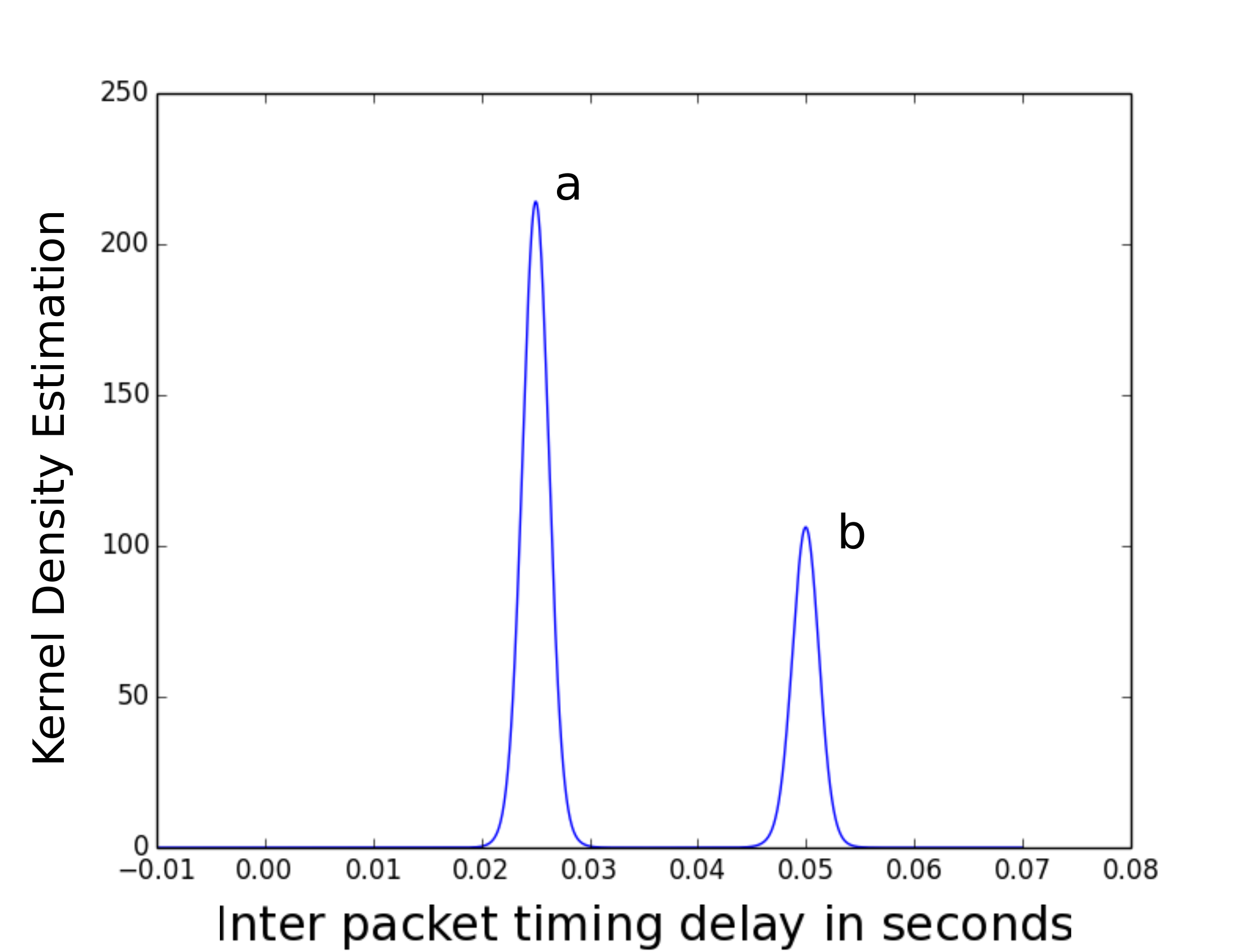}
\caption{The symbolization of inter-packet timing delays of legitimate PMU traffic}
\label{peakslegitimate}
\end{center}
\end{figure}

\begin{figure}[h]
\begin{center}
\includegraphics[width=1.8in]{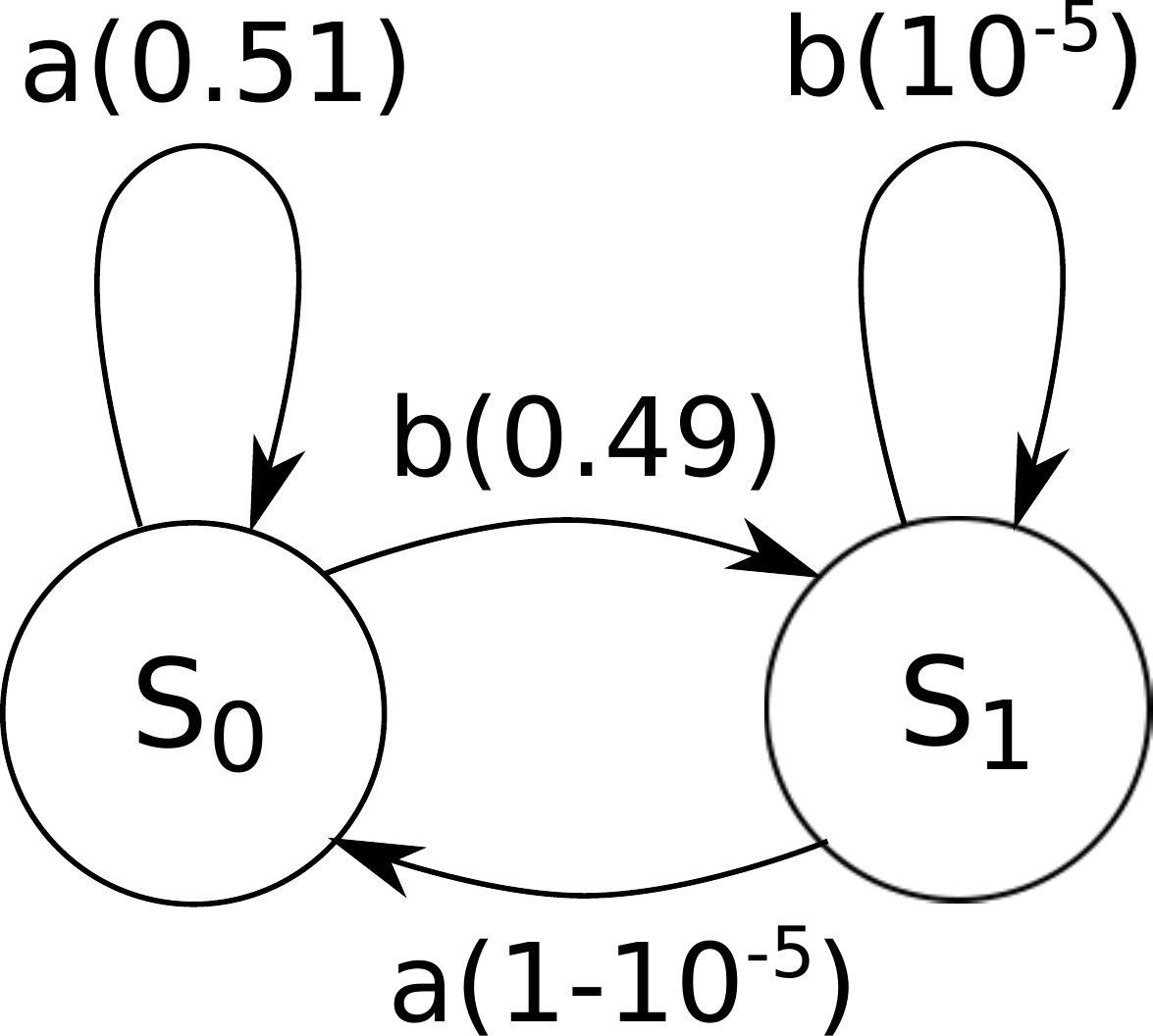}
\caption{The HMM of inter-packet timing delay side-channel of legitimate PMU traffic}
\label{HMMlegitimate}
\end{center}
\end{figure}

To infer the HMM representing the timing pattern of PMU traffic, packet collection is performed on the PMU side.  The time difference, $\bigtriangleup t$, is calculated by subtracting the receive time of the previous packet from the time of the current packet.  In other words, $\bigtriangleup t_i = t_i - t_{i−1}$ where $t_i$ is the receive time of packet $i$.  Obviously, the first packet in the data sequence will not have a value, so the process start with $i = 2$. This step is done so that network latencies between the source and the destination are
filtered out.  Using the calculated values of $\bigtriangleup t$,  the data is symbolized by grouping it into ranges and assigning anything in that range a unique symbol such as $a$ or $b$.  The symbolization result of is shown as Figure \ref{peakslegitimate}.  Given the symbolized data sequence, the zero knowledge HMM inference algorithm introduced in~\cite{schwier2009zero,yu2013inferring,chenmetric} is used to create the deterministic HMM. The inferred model is shown in Figure~\ref{HMMlegitimate}.


Given the model, the counterfeit PMU starts the timing SCM process by randomly selecting a start state in the model. To send the packet, a transition is taken from the current state and the corresponding delay is waited before sending a packet to the PDC. If there are more than one possible transitions out of current state, the transition is chosen randomly, weighed on the probability of each transition.

\section{Experiment}\label{sec:exp}
 \subsection{Traffic Collection}
\begin{figure}
\begin{center}
\includegraphics[width=3.5in]{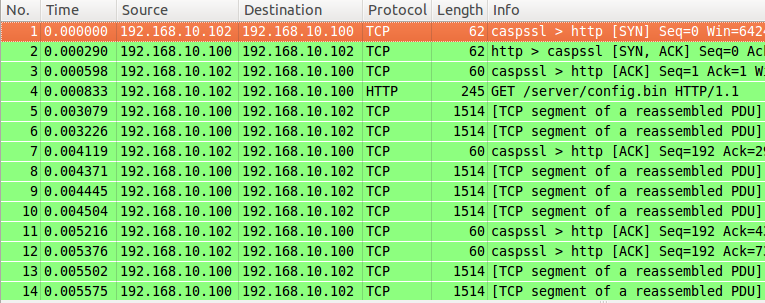}
\caption{Screenshot of botnet traffic (part)}
\label{bot-traffic}
\end{center}
\end{figure}

A Zeus C\&C server and a Zeus bot are set up in the lab on two Windows XP virtual machines. Figure \ref{bot-traffic} shows example botnet traffic. There are two protocols used: HTTP is for communication with the C\&C PHP server, and TCP for information exchange between bot and botmaster. The 
volume of the collected bot traffic is 1.4MB.

In the Real-Time Power and Intelligent Systems (RTPIS) Laboratory of Clemson University \cite{RTPIS}, a real-time power system is simulated with one hardware PMU and software openPDC (as PDC). Network traffic between PMU and PDC are collected with Wireshark on the same machine (windows 7 OS) running openPDC software. The traffic collection lasts around one hour, and the volume of collected traffic is 108.9MB.

\subsection{Experiment Setup}

\begin{figure}[!htb]
\begin{center}
\includegraphics[width=3.5in]{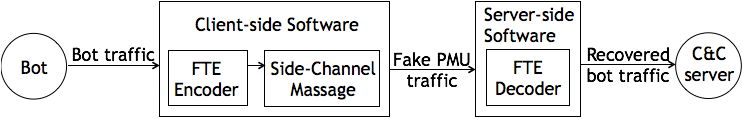}
\caption{Experiment setup.}
\label{setup}
\end{center}
\end{figure}

The experiment setup is in Figure~\ref{setup}. The implementation includes a client-side program (counterfeit PMU program) deployed on the bot and a decoder integrated as a front end of C\&C server. The client-side program takes the Zbot data stream as input, processes it with FTE then SCM.  It then sends the false PMU data over the network. When the data arrives at the server side, it is decoded by the FTE decoder and forwarded the to C\&C server. 



 \subsection{Experiment, Results and Analysis}





It is found that the bot and its C\&C server communicate fluently with the deployment of the traffic camouflage system.  This indicates the false PMU data is decoded correctly before forwarded to the C\&C server.  

\begin{figure}[h]
\centering
\begin{subfigure}{.5\textwidth}
\centering
\includegraphics[width=3.5in]{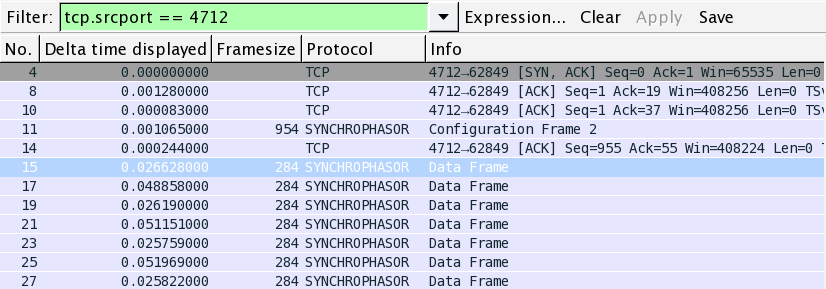}
\caption{Screen-shot of false PMU traffic (part)}
\label{botPMU}
\end{subfigure}
\\
\begin{subfigure}{.5\textwidth}
\centering
\includegraphics[width=3.5in]{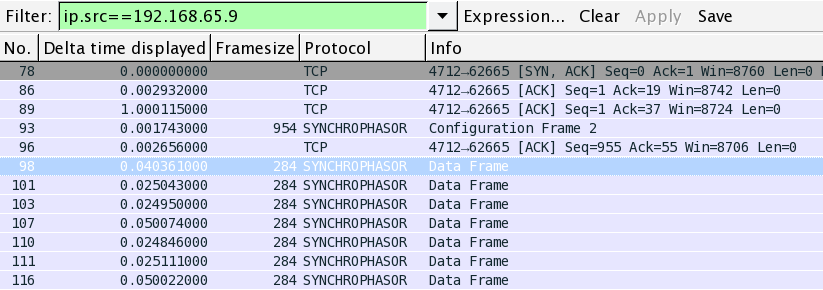}
\caption{Screen-shot of real PMU to PDC traffic (part)}
\label{realPMU}
\end{subfigure}
\caption{Comparison between false PMU traffic and real PMU to PDC traffic}
\label{fig:test}
\end{figure}

\begin{figure}[!htb]
\centering
\begin{subfigure}{.5\textwidth}
\centering
\includegraphics[width=3.5in]{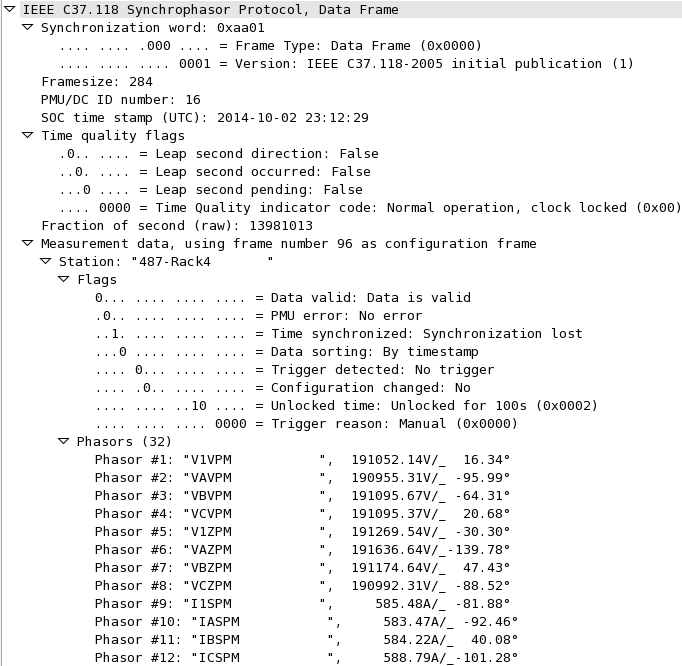}
\caption{Screen-shot of a typical camouflaged Zbot packet (part)}
\label{botPMUpack}
\end{subfigure}
\\
\begin{subfigure}{.5\textwidth}
\centering
\includegraphics[width=3.5in]{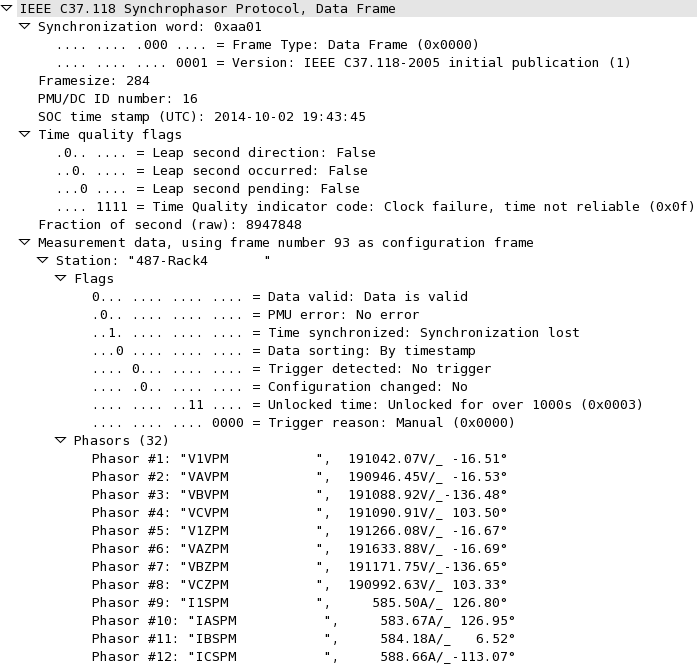}
\caption{Screen-shot of a typical PMU measurement packet (part)}
\label{realPMUpack}
\end{subfigure}
\caption{Comparison between camouflaged Zbot packet and PMU measurement packet}
\label{fig:test}
\end{figure}

As shown in Figure~\ref{botPMU}, the false PMU packets are recognized as the synchrophasor protocol, i.e., the protocol used for PMU-PDC communication.  The packet details of false PMU displayed in Wireshark are shown in Figure~\ref{botPMUpack}, which contain all the fields of a standard PMU packet as shown in Table~\ref{pmu-value}.  Values of each field are legitimate because they are consistent with the real PMU packets.  In particular, the voltage and current measurements are all within a reasonable range compared with real PMU packet shown in Figure\ref{realPMUpack}.


To further evaluate the proposed approach, side-channel analysis in~\cite{lu2011botnet} is launched against the false PMU data flow.  The confidence interval (CI) approach in~\cite{brooks2009behavior}  is used to determine if the observed false PMU traffic matches the PMU HMM (see Figure~\ref{HMMlegitimate}). 
The inter-packet delays of the false PMU packets are computed and symbolized.  Given the symbolized string, the CI method traces the data back through the HMM.  Meanwhile, transition probabilities and corresponding confidence intervals are estimated by frequency counting.  This process maps the observation data into the HMM structure.  After the confidence intervals are estimated, the percent of transition probabilities from originally given HMM that fall into their respective confidence interval are determined.  If this percentage is greater than a threshold value, it is accepted that the observations adequately match the HMM, i.e., the collected PMU traffic is legitimate in terms of timing. 

Receiver operating characteristic (ROC) curves are usually used to find the optimal threshold value.  By varying the threshold from 0\% to 100\%, the criteria for acceptance is progressively increased.  Since the HMM (see Figure~\ref{HMMlegitimate}) in this experiment only contains 2 states and each state has one probability distribution, there are only 3 possible values for the percentage of probability distributions that match, which are 0, 50\%, 100\%.   So there is no need for a ROC curve.  The detection results using these three values are given in Table~\ref{TPR-FPR}, where TPR denoted true positive rate, FPR denotes the false positive rate, and $l$ is the threshold used for detection.  TPR means the percentage of real PMU traffic that is correctly identified by Wireshark, while FPR means the percentage of false PMU traffic that is identified as PMU traffic.  As shown in the results, it is impossible to obtain a high TPR and low FPR at the same time no matter how $l$ varies. Therefore, it is difficult to perform an effective side-channel analysis.



\begin{table}[h]
\begin{center}
\caption{TPR-FPR with different thresholds }
\label{TPR-FPR}
\begin{tabular}{| c | c | c | c |}
\hline Threshold $l$ & $l = 0$ & $0 < l \leq 0.5$ & $0.5 < l \leq 1$ \\
\hline TPR & $1$ & $1$ & $0.0193$ \\
\hline FPR & $1$ & $0.9611$ & $0.0352$ \\
\hline \end{tabular}
\end{center}
\end{table}

A communication channel is built between the counterfeit PMU and software OpenPDC. All false PMU traffic is accepted by OpenPDC. This shows that the proposed method is capable of producing false PMU data stream close to real PMU.





\section{Conclusion and Future Work}\label{sec:candf}
In this study, one protocol is transformed into another to disguise the original traffic and deceive existing network traffic detection approaches, including side-channel analysis. As an illustration, Zbot C\&C traffic is transformed to PMU data stream.  The experiment shows that the syntax of the counterfeit PMU packet is consistent with real PMU traffic.  The proposed method is also resistant to timing side-channel analysis. False PMU packets are accepted by OpenPDC without any errors.  Other target protocols can easily be used.


However, the proposed traffic camouflage counter-countermeasure is not perfect. Future study includes: (1) The current camouflage is off-line.  On-line implementation is needed. (2) The throughput can be increased by determining the optimal number of character positions used to represent a packet field. (3) Methods of selecting appropriate target protocols are needed. (4) The ultimate goal of this study is to make anti-virus community aware of this new way of hiding malware traffic and therefore effective countermeasures are needed.

\bibliographystyle{IEEEtran}
\bibliography{mybib}

\end{document}